# On the benefit of 3-tier SOA architecture promoting information sharing among TMS systems and Brazilian e-Government Web Services: A CT-e case study

Thiago Suzuki[a], Larissa Romualdo-Suzuki[b]

[a]*Fundacao Getulio Vargas, Ribeirao Preto, Brazil (thiagosatoshi@gmail.com)*
[b]*University of Sao Paulo, Department of Electrical Engineering, Sao Carlos, Brazil*

**Abstract**

Technological advances regarding software integration processes have revolutionized the communication between government and society, which are increasingly based on information and communication technologies (ICT's). Service-Oriented Architecture (SOA) has emerged originating new prospects for system integration within organizations and external partners, providing essential information for decision-making process. Brazilian e-Government initiatives has introduced as a national electronic document model known as Electronic Transportation Knowledge (CT-e), looking for simplifying additional obligations of taxpayers and, at the same time, allowing real-time monitoring of cargo transportation services provided by the Revenue. Nonetheless, there is a major challenge that prevents Transportation Management Systems (TMS) and Government to be benefited by this innovation, due to their distinct platforms and databases that prevent information exchange between them. This paper proposes an architectural solution to integrate TMS systems with CT-e Web Service applying concepts of SOA and N-tier architecture. Through a real case study of a large Cargo carrier, we report an increase in transportation knowledge management, speeding up in the communication and data validation process and several costs reduction including paper, printing, document storage and those involved in the necessary logistics that is needed to recover such documents.

## 1. Contextualization

The notorious technological advances in electronics has brought tremendous impact on the ways society communicates, resulting in the discovery of the intrinsic relationship that exist between the digital computer and digital communication. As soon as the differences were exposed, the similarities became so prominent that a marriage between these two was imminent. This happening implied such turmoil in the analogue communication media that almost all became immediately digital, as fruits of that "almost" perfect wedding. Consequently, it revolutionized the communication among people, society, nations and so for and, especially the internet infrastructure, have brought new models for relationship between government and society, which are increasingly based on information and communication technologies (ICT's). Therefore, digital communication technologies play, nowadays, a crucial role in how governments fulfil their main duties, allowing that the public administration becomes more efficient, more democratic and more transparent. The first e-Government initiatives have occurred in the late 90's but, until recently, Governments have launched digital electronic projects in order to provide digital services and information to their citizens and businesses [1]. E-Government practices comprises the use of ICT's, in particular the Internet, to provide public information, to offer services to all participants, to improve internal processes and to integrate the interactions and interrelationships between government and society [2], [3], [4], [5], [6]. Hence, e-government when properly executed can be an important key regarding institutional reforms, the provision of goods and services to the public sector and in government procurement.

In particular, it is known that the Brazilian e-Government initiatives, have started in the late 90's and featured a major advance in early 2001, by offering more than 800 types of electronic services [7]. Most Brazilian e-Government initiatives have focused on providing web-based services to their citizens, Government-to-Citizen (G2C), and enterprises, Government-to-Business (G2B) [8]. The possibility for designing new digital projects along with the elimination of fiscal documents arose with the Digital Certification established in 2001, becoming a further step towards advances in technologies between revenue and taxpayers (G2B). In summary, electronic tax documents simplify the form by which taxpayers fulfil their obligations as well as allow better monitoring of commercial operations by the tax authorities, proving to be a beneficial solution to those involved in transactions with these documents. Tax administrators face the challenge of adapting to the process of globalization and digitalization of commerce and transactions between taxpayers. The transactions volume and the amount of handled resources increase in an intense pace and, in the same proportion, increase the costs to detect and prevent tax evasion (Internal Revenue Service). This advance was the first step in searching for solutions to improve administrative integration, standardization and to increase quality information, streamline costs and workload in operational service. It allows also, more effectiveness in the supervision, greater possibility of carrying out coordinated and integrated tax actions, increase the possibility of exchanging tax information among the various governmental levels, cross-scale data with standardized data and uniform procedures.

In 2006, the Electronic Transportation Knowledge (CT-e) project was introduced as a national electronic document model to replace the current system that issues tax documents on paper, with legal validity guaranteed by the digital signature of the issuing company [9]. This project seeks to simplify additional obligations of taxpayers and, at the same time, allow real-time monitoring of cargo transportation services provided by the Revenue. It is expected to reduce costs of shipping and storage of tax documents, time that trucks are required to spend at Fiscal border posts, encouraging the use of electronic relationships with customers (B2B), in addition to the standardization of electronic relationships among companies. Furthermore, the benefits of this initiative go beyond the ones provided to the Government and Cargo carriers, yet it reflects benefits to the society as a whole in reducing costs of printing and paper purchasing, helping environmental preservation.

Nonetheless, there is a major challenge that prevents Cargo carriers to delight in all these benefits and innovations. TMS systems which are the CT-e issuers have different platforms and databases in comparison with the ones used by the CT-e project XML Web Service. Therefore, the information exchange between these systems requires good expertise and effort from system engineers. This paper proposes an architectural solution to integrate TMS systems with CT-e Web Service by applying concepts of SOA (Services Oriented Architecture) and N-tier architecture.

Through a real case study of integration between a large Cargo carrier and the CT-e Web Service, results show an increase in the transportation knowledge management, speed up in the communication and data validation process and several cost reduction. Beyond that, the cost reduction comprises not only in paper, printing and document storage, but also the cost involved in the necessary logistics that is needed to recover such documents. This paper is organized as follows. Section 2 provides a general introductory background to the CT-e architecture and its communication design. In section 3 is shown a brief description of SOA and N-tier architecture, as well the proposed integration framework. Section 4 presents an analysis of the pointed results and the concluding remarks.

## 2. CT-e Architecture

The CT-e can be conceptualized as an exclusively digital document that is electronically issued and stored in order to document each provision of transport service (by highway, aerial, pipeline, waterway or rail), whose validity is guaranteed by the issuer's digital signature and its respective authorization provided by the Revenue located in the taxpayer's State of domicile [9]. Basically, as the CT-e is a stand-alone document it must to be issued one by one and, each CT-e must to have its own digital signature. Nevertheless, its transmission process must to be performed in batches, in which it is possible to encompass up to 50 CT-e (i.e., it can encompasses even a single CT-e) and should not, however, exceed the maximum size of 500 Kbytes. Information exchanging of batches between the TMS system and the CT-e Web Services should be performed via XML Schema. The adopted XML document specification follows W3C recommendation for XML 1.0 [10], in addition to UTF-8 character encoding. Moreover, this electronic document must be digitally signed with a digital certificate with contains the CNPJ (registration at the Brazilian National Registry of Legal Entities) of the head office or the CNPJ of the establishment issuing the CT-e request. A SSL protocol validates this digital certificate when it is transmitted to CT-e Web Service. Figure 2.1 illustrates the concept of CT-e and batches of CT-e and their basic XML structure (for illustration purposes only as the real XML containing CT-e batch is much more complex than this).

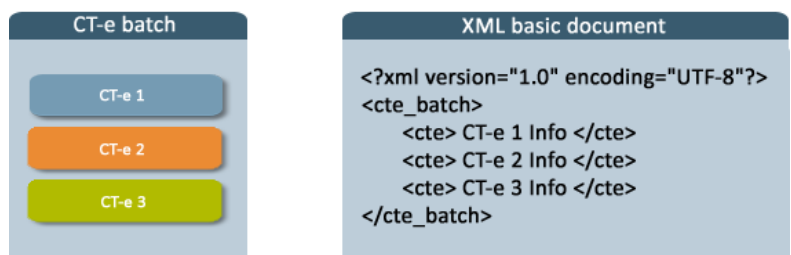

**Figure 2.1. Illustration of a CT-e batch containing 5 CT-es and its respective basic XML structure.**

CT-e project provides seven online services and they are shown in Table 1. For each requested service there will be a specific Web Service. Communication flow is always originated from the taxpayer system by sending a request message to the Web Service stating the desired service. The provided services can be either synchronous or asynchronous depending on the requested service processing and on the Web Service demand. Therefore, the requested service can be answered at the same time of the received request or stored in a queue of critical processing services for a better performance of the CT-e project's communication resources.

**Table 1**
Services provided by the CT-e Web Services.

| Services | Type | Description | Input |
| --- | --- | --- | --- |
| Sending CT-e Batch | Asynchronous | Service available to obtain CT-e batches | XML structure comprising transportation knowledge batch |
| Tracking CT-e Batch Processing | Asynchronous | Service available to provide CT-e batch processing result | XML structure comprising CT-e receipt number |
| Withdrawing CT-e | Synchronous | Service available to withdraw CT-e | XML structure comprising withdrawing message and its respective request approval |
| Withdrawing CT-e numbering | Synchronous | Service available to withdraw CT-e numbering | XML structure comprising numbering withdrawal message and its respective request approval |
| Tracking CT-e current status | Synchronous | Service available for CT-e status tracking | XML structure comprising CT-e access key |
| Correcting CT-e Batch | Synchronous | Service available to CT-e batch rectification | XML structure comprising CT-e rectification |
| Tracking service status | Synchronous | Service available to track service status | XML structure comprising service status checking |

In synchronous service, the requested service is processed and concluded on the same connection and the taxpayer receives a message containing the processing result of the requested service. In contrast to this, in asynchronous service the processing of the request is not performed on the same connection, nonetheless the taxpayer receives an acknowledgement receipt of the requested service. In order to process services requests in the queue, the CT-e system will remove the request from the entry queue accordingly to the order of arrival (FIFO) and then store the processing result on an outgoing queue. Figure 2.2 illustrates synchronous and asynchronous services available on the CT-e platform.

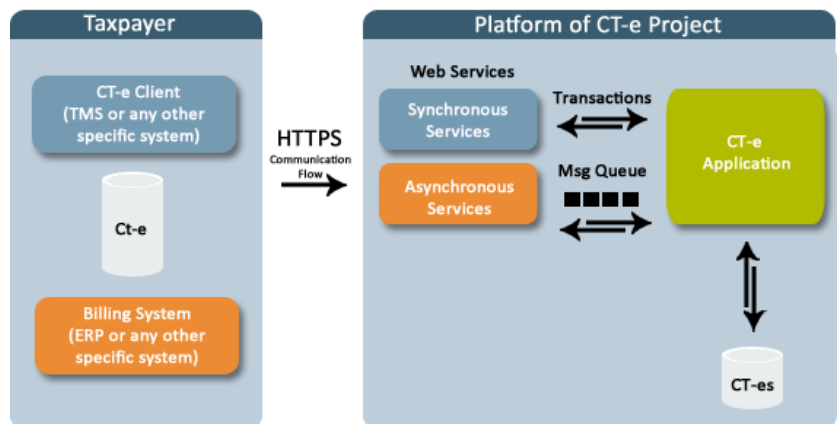

**Figure 2.2. CT-e architecture – synchronous and asynchronous services. Adapted from (CT-e, 2011).**

The average response time which measures the performance of batch processing service is calculated based on the time elapsed between the time of receiving the message and the time of the request processing result storage in the outgoing queue. There is a limit of 50 CT-e batch per connection. The grouping of all CT-e within the batch must be done, by an operational and control restriction, respecting the rule that all CT-e batch must be from the same establishment. As early stated, the maximum size of the CT-e batch is limited to 500 Kbytes, therefore the taxpayer must create a CT-e batch that does not exceed this limit, even though the amount of CT-e batch is within the limit of 50 CT-e. The processing result of *Sending CT-e Batch* service requisition, asynchronous service, should be available in the output queue for a minimum time of 24 hours. Any identified error incidence in the validation of the received data will entail a process interruption and a message containing both error code and its description will be available. In successful cases a receipt number will be returned containing date, time, receiving place and average service time response in the last 5 minutes. CT-e also encompasses the printing of a paper-based document known as DACTE (Electronic transportation knowledge auxiliary document), whose function is monitoring provided services, and consequently, delivery of goods, and also to facilitate CT-e tracking on the Internet. Although there is still paper-based issuing document, however, this document can be printed on A4 plain paper instead of three copies that was needed to print on a specific invoice form before this service be available.

## 3. Software Engineering Architectures

*3.1 Introduction*

Software Engineering intends to provide a high quality software product in which organizations combine their business and components to provide integrated products to their end-users. In the early days of computing, software development was limited to monolithic applications, which encompasses all functionality in a single tier. Due to this, its maintenance and updating was extremely hard and complex. In order to reduce redundancy and inconsistency due to the use of non-centralized repositories, this early technology evolved to two-tier applications, where database access was separated from the main application (client / server), consequently allowing database sharing between various systems [11].

Basically, the first layer is responsible for the presentation data to the user and the second layer is responsible to provide data services to the client. Typically, a client runs on end-user desktops and through a network it interacts with a centralized database server. This architecture provides many advantages, for instance, enables wider access to databases on the network, increases performance as client and server are able to process parallel applications, reducing in hardware and communication costs and increases data consistency. Despite this advance, this new architecture was still monolithic due to presentation layer (human-machine interface - HMI) and business logic layer was grouped into a single tier.

Moreover, as applications complexities are increasing over time, especially regarding web-based applications evolvement, the need for enterprise scalability has changed the two-tier architecture scenario by introducing an additional tier [12]. Thereupon, three-tier architectures had emerged splitting systems functionalities into a presentation layer (*client*), business logic layer (*application server*) and data logic access layer (*database server*). The presentation layer provides to the end-user the system interface, business logic layer encompasses all business rules and data processing and, data logic access layer stores the data required by the business layer. This new architecture provides many advantages over the last two architectures, for instance, it is more flexible by allowing tiers being changed and amended independently, reduces hardware costs as the *client* is 'thin' and software distribution issues. Furthermore, any changes in a layer do not affect others, as long as their communication mechanism remains unchanged. In order to provide more flexibility and scalability, the three-tier architecture explained in section can be further expanded to n-tiers. For instance, between the application and business layer a Web service layer could be added to support interoperable machine-to-machine interaction over a network [12]. The concepts of Web services are explained in the following section. We can adapt a 3-tier architecture incorporating a Web services that integrate and provide systems accordingly to their specification. Figure 3.1illustrates the client/server, two-tier, three-tier and n-tier architectures.

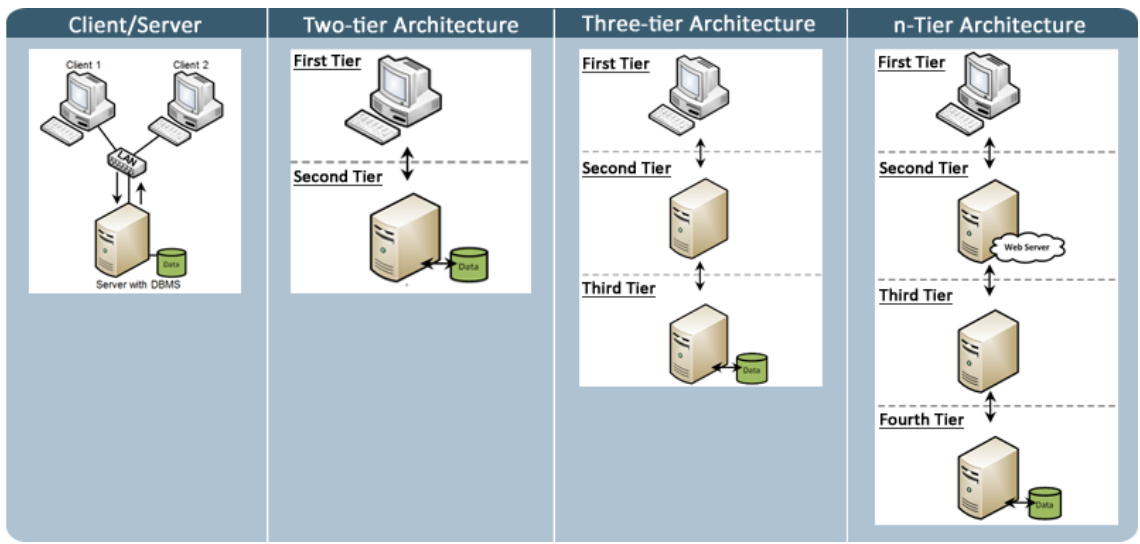

**Figure 3.1.Tiers client/server architecture.**

*3.2 Software Integration*

Frequently, there are an extensive number of coexisting systems and information repositories within a company that are developed to meet specific requirements of each department. Thereby, those systems could have been developed either in distinct languages, platforms, databases or source of information coming from external sources. As a consequence of this development, for instance, there is almost no reuse of solutions, interoperability issues and maintenance should to be carried out separately for each system. Moreover, a very high time-demand is required in the development stage and, due to data scattering in non-centralized repositories, often there is redundancy and inconsistency on the stored data. Thereupon, it is a hard task to obtain reliable and consolidated information coming from a wide range of software systems within an organization [13]. Therefore, this is not the ideal architecture as each client must have their unique architecture to be able to exchange information within servers. Software Integration emerges as a suitable solution in such environments. This solution allows information sharing within an organization or with external partners, providing essential information for decision-making process [14].

The main purpose of system integration is to achieve systems that can easily manage data and procedures access without any obstacle. Hence, this concept facilitates the implementation of joint activities between distinct systems and to allow organizations to be set to attend constant demand and changes in their environment. The architectural change when using integration architecture allows linking all participants, i.e., clients and server into a unified way. The evolution on integration technologies has occurred as a response to the need of flexible approaches. Nowadays there are several valid and different systems integration approaches for the same problem, nevertheless the choice of the best approach must to be based on the business and technological reality of the organization.

*3.1 Web Services*

In the past decades, an uncountable number of businesses all over the world have been adopting the World Wide Web (WWW). Recently, these distributed systems have improved the integration among web applications [12] originating the concept of Web Services. The main purpose of Web Services is to overcome Web integration constrains such as lack of common interface, information sharing, independence on operational system and hardware and, to serve as basis for the new generation of B2B and EAI applications. Therefore, Web Services allow better communication between systems regardless system's programming language and platforms. Although it provides a most needed service as the communication between different applications geographically distributed, a key issue for the success this emerging technology is to offer adequate support to Service of Quality. Thus, this technology will provide high quality and continuous delivery of services [16].

Web services encompass the use of well-known techniques and standards, such as XML, SOAP, WSDL and UUDI. To achieve a successful connection to a Web Services, firstly it is required locating it, finding the interface with its semantic call, in addition to adjust systems to service collaboration. The interface containing the Web Service's semantic call, describes messages that it is able to understand, restrictions applied to data within them, their categories or ontologies, in addition to descriptions on how join a Web Service forces with another one to support business workflow. This semantic for Web Services interfaces are manipulated by Web Services Description Language (WSDL), which corresponds to a XML (*eXtensible Markup Language*) document that describes the specification of a service and the way messages are exchanged between clients and service providers [17]. This semantic also describes a set of SOAP (*Simple Object Access Protocol*) messages, which is a XML-based communication protocol and is typically the responsible for the interactions between Web Services. SOAP provides support to document-based systems, where message is just a XML document wrapper. XML-Schema has been widely used as a technique to define complex data types into XML documents [12]. The UDDI (*Universal Discovery, Description and Integration*) protocol is a platform-independent XLM-based registry that is responsible to answer SOAP messages, providing access to WSDL documents which contains all requirements needed to interact with the Web services controlled by this registry. All this are the foundation to a new generation of platform-independent distributed systems.

*3.2 Service-Oriented Architecture*

Due to the extensive adoption of Web Services, Service-Oriented Architecture (SOA) has emerged as a solution that allows the communication between systems originated in distinct programming language and platform. It focuses on integrated structuring of business activities, sharing and reusing services in distributed environments rather than on development of isolated solutions [18], [19], [20]. In business context, its functions are defined by means of software components defined as services and, these services are available over the network to a wide range of systems from diverse organizations. Thereupon, customers, services providers and records goals to invoke, locate, connect and dynamically execute services using standard languages such as XML [10], [12], [21], [22].

An example extracted from [12] illustrates three processes: Service Scheduling, Order Processing and Account Management (Figure 3.5). Each process is accessing six distinct databases. It can be noticed that there are common "services" that is performed by all this processes. Any adjustment in business requirements will require an extensible amount of time to adapt these processes. SOA architecture is a very promising solution to overcome this constrains due to its ability to provide services independently of applications and platform. Figure 3.5 illustrates a traditional IT achitecture for three business process and the same architecture after introducing SOA architecture. It is clear that SOA provides many benefits for IT architecture as it allows reuse of business services, providing a collaborative, interoperable and integrated architecture.

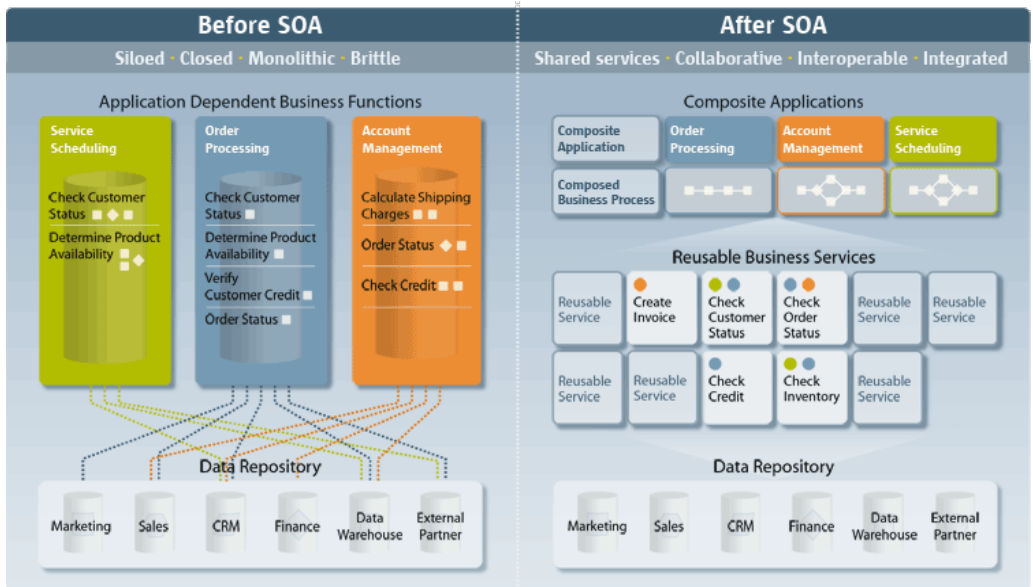

**Figure 3.5.** (a) Traditional IT architecture for three business processes; (b) service-oriented architecture that splits the processes into a number of reusable services. Adapted from (Connolly & Begg, 2010).

## 4. Materials and Methods

*4.1 Introduction*

In this work it is proposed the implementation of an integration system solution based on SOA and N-tier concepts to integrate TMS systems and the Web Services of CT-e project of the Brazilian Federal Government (SEFAZ). The TMS system is based on COBOL programming language and is recently being replaced by a .NET-based system. It was used the *Microsoft Visual Studio 2008* using C#.NET programing language and PL/SQL to allow the connection with the ORACLE 10g database. In this work it was developed three front-end clients: a system to manage all CT-e issued by the Cargo carrier organization, a Windows service that is responsible to exchange information between the TMS system and CT-e Web Services and, a system to obtain information to generate DACTE document file. Before parameterizing the integration system it is required from Cargo carrier to be registered with SEFAZ, to have the digital certificate and a password issued by SEFAZ.

Essentially, before Brazilian e-Government launch CT-e project TMS systems were issuing paper-based document that allowed goods delivery. However, with this approach, sometimes there was data inconsistency on this document and Government had no control over tax evasion. When this project was applied in practice, for the TMS system reported in this work, was not easy to change all system platform and logic to attend this new reality. Therefore, the easiest way found by this Cargo carrier was to adopt an integration system, where TMS system issues a TXT file containing all CT-e information that was used to generate the paper-based document. This TXT file is stored into folder **IN** and all results obtained for TMS system's solicitation is stored into folder **OUT** in TXT file format. Figure 4.1 illustrates the integration architecture developed in this work. The integration architecture and all use cases of the integration system are explained in the following section.

*4.2 Integration Software Architecture*

*4.2.1 Presentation Layer*
The **Windows Srv. App** is a windows service application developed to take responsibility for reading and processing all documents issued by the TMS system and sends them to the Web Service containing the business tier, which will generate a proper XML file containing the CT-e to be transmitted to CT-e Web Services. Each 1 minute, this windows service application checks if there is an available TXT document to be converted into a CT-e batch file. In a positive case, this application generates each CT-e batch containing up to 50 transportation knowledge and 500k. After this stage, the all CT-e batch is transmitted to the CT-e Web Services. In case of a synchronous service, the CT-e batch is processed on time and the windows service application get and originates to the TMS system a file containing the CT-e processing results. Otherwise, in case of an asynchronous service, the windows service application consults every minute the status of the transmitted CT-e batch until it receives the result of the CT-e processing. Two additional client systems were developed to improve CT-e management that is the **CT-e Mgmt. App**, which allows Cargo carrier's administrators to check CT-e status, error in CT-e validations, in addition to verify the quantity of issued CT-e and, **Printer App** is responsible to obtain information to generate DACTE document.

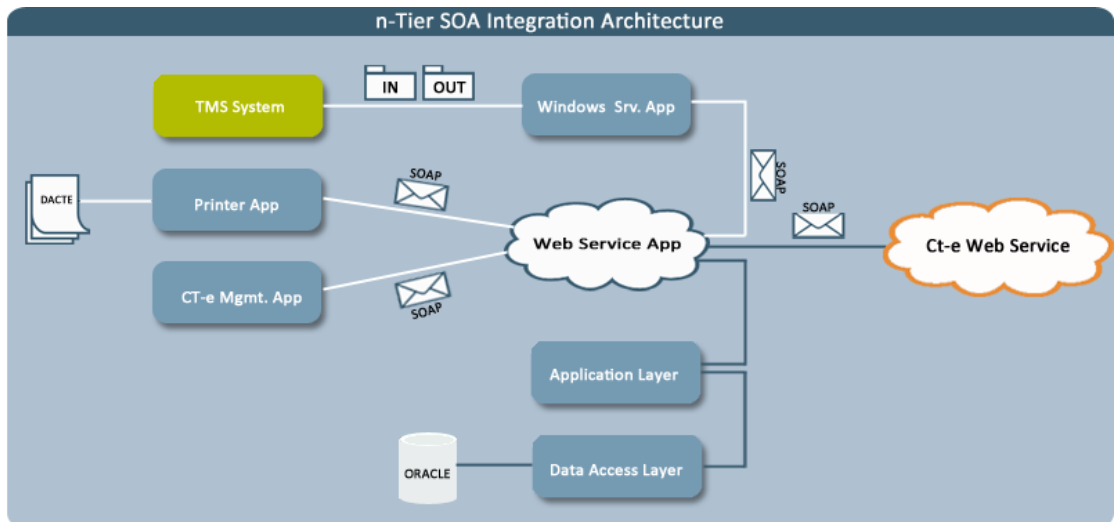

**Figure 4.1. 3-tier SOA architecture for TMS system integration with CT-e Web Services.**

*4.2.2 Web Service App*
As stated before, CT-e project is a new initiative in Brazil and hence, some changes are being made to improve the project and overcome any particular constraint. Given this, system's integration business and data access tiers were moved from local to a Web Service, referred here as **Web Service App**, making easier any changes in the business logic layer and allowing these changes to be applied remotely not only to this Cargo carrier, but to other further clients since the transmission of CT-e is going to be mandatory in few years.

*4.2.3 Application Layer*
In this tier, it was implemented all business logic required to generate a proper XML file containing the CT-e to be transmitted to CT-e Web Services, refer to batch processing, withdrawing CT-e, withdrawing CT-e numbering, tracking CT-e current status, correcting CT-e batch, request to data access tier to insert, delete and update all CT-e, tracking service status and further logics for management purposes.

*4.2.4 Data Access Layer*
This tier is responsible to establish connection with the database and contains all SQL transactions such as insert, delete, update and selects. The Cargo carrier studied in this work has provided an ORACLE 10g database that allows storage of all CT-e information, which encompasses CT-e data, CT-e batch, withdrawing information among other relevant information. In this database is stored a

# 5. Results

Through conducting extensive software requirements survey and studying client's IT environment, it was possible to develop a very suitable integration technique that allows TMS system to exchange information between Brazilian e-Government Web Services (CT-e). CT-e project provides two environments for CT-e receiving. The "approval" environment is specific to the testing and integration of taxpayer applications during the implementation and compliance of the taxpayer CT-e emission system. The issue of CT-e in the "production" environment is subjected to the approval of the taxpayer's IT and business staff, which should assess the adequacy, behavior and performance of the CT-e issuing system in the approval environment.

Once approved the CT-es in the "approval" environment, taxpayer shall be enabled to the production environment (CTE, 2011). To validate the proposed architecture, for three weeks CT-es were issued and transmitted to Web Services in the approval environment. All results were analyzed whether CT-es were either accepted or rejected (e.g. due to system misconfiguration or user mistyping). In our experiment it was noticed that the most common error occurred in the CT-e batch was due to CNPJ mistyping. The studied Cargo carrier provides services for about 2.000 cities and encompasses nine States in Brazil. Daily, this Cargo carrier issues about 1.000 CT-es and transmits them through integration system over the Internet to the CT-e Web Services. Within two minutes, a CT-e batch was already processed by CT-e application, all CT-e within the batch was tracked and the DACTE documents were printed.

Results show that this architecture was able to proper integrate TMS system with CT-Web Services, provided a flexible and manageable integration solution, increased transportation knowledge management through the **CT-e Mgmt. App**, and speeded up communication and data validation process. Further benefits were also identified as several costs reduction that encompasses paper, printing, document storage and those involved in the necessary logistics that was needed to recover such documents.

## 6. Discussion and Conclusions

Through the study of a real data transmission between client and the Government's web service, it was noticed that the development of this type of application provides many benefits, since it integrates systems that would not be able to exchange any information, promote improvement in the knowledge transportation management, in addition to speeding up the process of communication and data validation.

The advantage in using the N-tier methodology is that the presentation logic is separated into its own logic and physical layer. The features of the business layer could be divided into classes and these classes were grouped into packages and components, reducing the dependencies between classes and packages. It allows its reuse in different parts of the application and even by other application modules. Indeed, findings reveal that the SOA architecture was the best solution to integrate TMS system and CT-e Web Services as CT-e project is a new initiative in Brazil and its Web Service architecture has been changed since it was launched. Therefore, using SOA architecture allowed changing business logic once and making available the new changes or features to all participants involved in this integration.

Another interesting finding was the cost reducing in tax documents storage as well as all needed logistics for their recovery. Before CT-e project was launched, the paper-based tax documents should be kept by taxpayers and submit them to tax authorities into preclusive deadline. Hypothetically, if a taxpayer daily issues 100 transportation knowledge, within a month there will be approximately 2.000 tax documents and about 120.000 at the end of five years. When sending documents electronically, the electronic document remains under the responsibility of the taxpayer and the costs involved in digital archiving is less expensive than the cost of the physical storage. With the electronic validation of all CT-e, information misspecification was dramatically reduced in the CT-e issuing by the Cargo carrier. Therefore, the integration architecture has enabled optimization in organization, storage and management of CT-es, facilitating the retrieval and exchange of information. Furthermore, the development of this application reflects benefits to the society as a whole in reducing costs of printing, paper purchasing, shipping, storage and logistics of tax documents, standardized and encouraged the use of electronic relationship among customers and organizations (B2B).

**Appendix A**

*A.1 UML modelling*
In this section is presented graphical descriptions of all system's use cases using activity and system sequence diagrams.

*A.2* Sending CT-e Batch
This use case allows the Windows App, who is responsible to send CT-e data to the Web Service, to obtain data from the TMS system and forward this data to CT-e Web Services. This is an asynchronous service provided by the CT-e Web Service responsible to receive CT-e batches. As preconditions, it is required from Cargo carrier to be registered with SEFAZ, to have the digital certificate and a password issued by SEFAZ. Table 2 presents textual description of this use case. Figure 4.2 illustrates the system sequence diagram.

**Table 1**
Textual description of Sending CT-e Batch use case.

| Sequences | Description |
| --- | --- |
| Main Success sequences | 1) TMS system generates a TXT file containing all CT-e to be sent to **Windows Srv. App.**<br>2) **Windows Srv. App** obtains CT-e and sends to **Web Service App.**<br>3) **Web Service App** inserts CT-e into a queue up to 50 CT-e at a limit of 500k and then build a XML which consists of the CT-e batch.<br>4) **Web Service App** inserts into ORACLE database produced CT-e batch.<br>5) **Web Service App** sends to CT-e Web Service each produced CT-e batch.<br>6) CT-e Web Service performs some initial validations (digital certificate, CT- batch size, XML formatting).<br>7) CT-e Web Service inserts each batch into an FIFO service-based input queue, generates a receipt number and forwards it to **Web Service App.**<br>8) **Web Service App** inserts each receipt number into ORACLE database on its respective batch register and later refer to it to get the processing result.<br>9) **Web Service App** ends its communication with CT-e Web Service.<br>10) As the CT-e batch queue decreases, CT-e Web Service transfers each batch to CT-e Applicative to be processed. |
| Error sequences | E1) Problem found in validation Digital Certificate (invalid, revoked, overdue, prerequisites violation, CNPJ).<br>E2) Problem found in validation XML file (size, formatting, CT-e Web Service status).<br>E3) Problem found in establishing connection with CT-e Web Service (invalid / absent Brazilian federation Unit, version issues).<br>*Error sequences start at point 6 of main success sequences.*<br>7. CT-e Web Service rejects CT-e batch and issues an error code. |

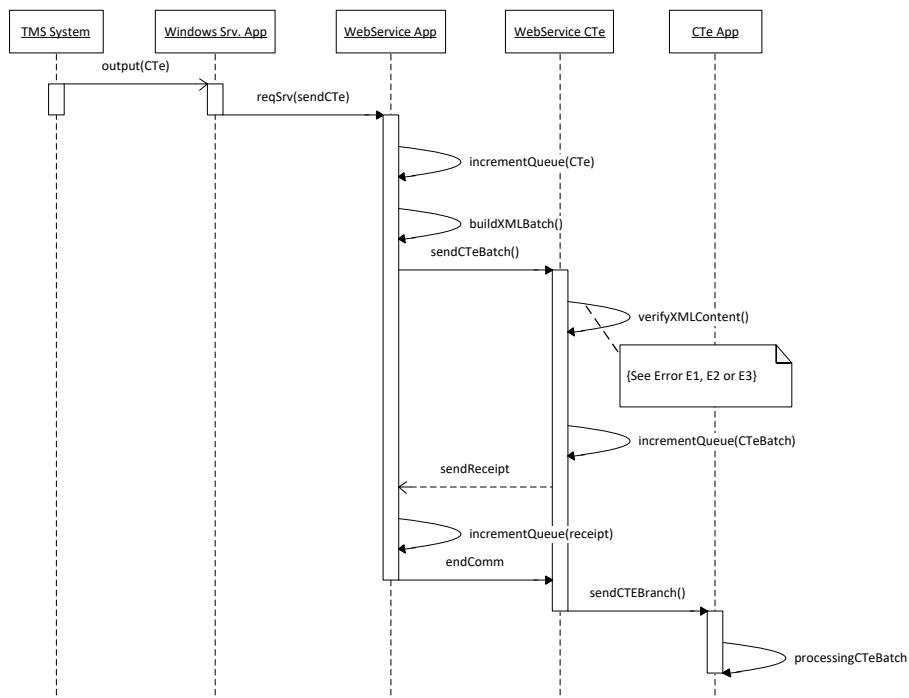

**Figure A.1 Detailed system sequence diagram of the *Sending CT-e Batch* main scenario.**

*A.2 Tracking CT-e Batch Processing*
This use case allows the Web Service App, who is responsible of tracking CT-e batch in the CT-e Web Service, to track CT-e batch in the CT-e Web Service and to obtain a XML file containing the processing result. This is an asynchronous service provided by the CT-e Web Service responsible to track CT-e batches. There are two possible processing results: processed or processing. As preconditions, it is required from Cargo carrier to be registered with SEFAZ, to have the digital certificate, a password issued by SEFAZ and a valid CT-e receipt number. Table 3 presents textual description of this use case. Figure 4.3 illustrates the system sequence diagram.

**Table 2**
Textual description of Tracking CT-e Batch Processing use case.

| Sequences | Description |
| --- | --- |
| Main Success sequences | 1) After processing the requested service, CT-e Applicative inserts each processed batch into an output queue containing the original XML data file containing the code of the processing result.<br>2) **Web Service App** obtains receipt number of all transmitted CT-e batch<br>3) **Web Service App** builds a XML document comprising CT-e batch receipt number and sends to CT-e Web Service.<br>5) CT-e Web Service performs some initial validations (digital certificate, CT- batch size, XML formatting).<br>6) CT-e Web Service queries and tracks the processing result.<br>7) ) CT-e Web Service issues a XML document containing the processing result.<br>7) **Web Service App** obtains the processing results and update in ORACLE database for each batch the "processing status" field with the received result (i.e, processed, processing).<br>8) **Web Service App** ends its communication with CT-e Web Service. |
| Error sequences | E1) Problem found in validation Digital Certificate (invalid, revoked, overdue, prerequisites violation, CNPJ).<br>E2) Problem found in validation XML file (size, formatting, CT-e Web Service status).<br>E3) Problem found in establishing connection with CT-e Web Service (invalid / absent Brazilian federation Unit, version issues).<br>E4) Problem found in validating receipt number (invalid / absent Brazilian federation Unit, invalid digital signature, invalid CNPJ, inconsistency in XML information).<br>*Error sequences start at point 5 of main success sequences.*<br>6. CT-e Web Service rejects CT-e batch and issues an error code. |

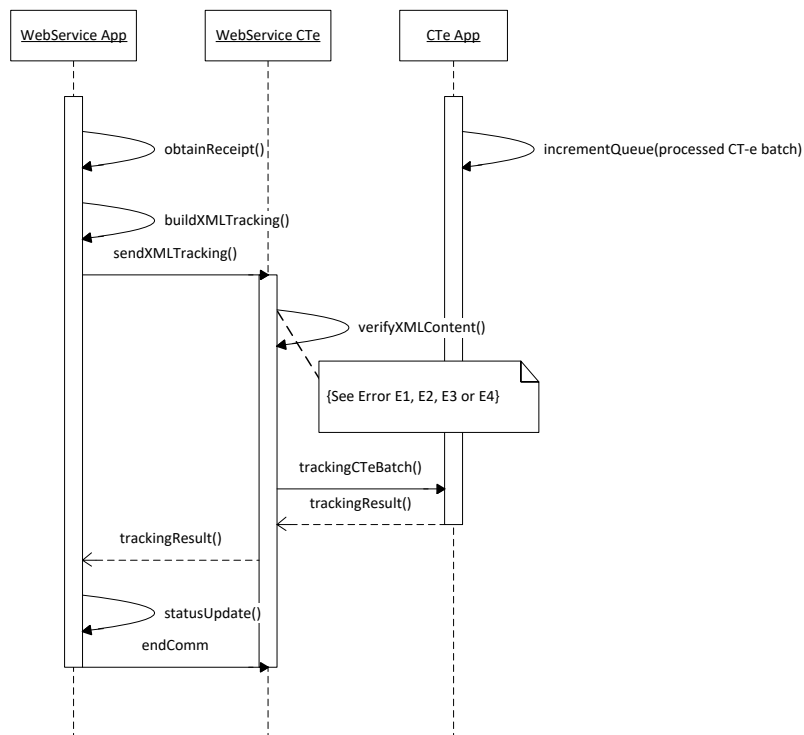

**Figure A.2. Detailed system sequence diagram of the *Tracking CT-e Batch Processing* main scenario.**

*A.3 Withdrawing CT-e*

This use case allows the Web Service App, who is responsible to send CT-e batch data to the CT-e Web Service, to obtain a XML document from the CT-e Web Service containing the withdrawing processing result. This is a synchronous service provided by the CT-e Web Service responsible to withdraw CT-e batches. As preconditions, it is required from Cargo carrier to be registered with SEFAZ, to have the digital certificate, a password issued by SEFAZ, a valid CT-e batch identifier (key). Table 4 presents textual description of this use case. Figure 4.4 illustrates the system sequence diagram.

**Table 3**
Textual description of Withdrawing CT-e use case.

| Sequences | Description |
|---|---|
| Main Success sequences | 1) TMS system generates a TXT file containing all CT-e to be withdrawn and sends it to **Windows Srv. App.** <br> 2) **Windows Srv. App** obtains CT-e and sends to **Web Service App.** <br> 3) **Web Service App** inserts CT-e into a queue up to 50 CT-e at a limit of 500k and then build a XML which consists of the CT-e batch. <br> 4) **Web Service App** updates into ORACLE database CT-e status to "cancelling". <br> 5) **Web Service App** sends to CT-e Web Service each produced CT-e batch. <br> 6) CT-e Web Service performs some initial validations (digital certificate, CT- batch size, XML formatting). <br> 7) CT-e Web Service sends withdrawing request to CT-e Application. <br> 8) CT-e Web Service issues a XML document containing the processing result. <br> 9) **Web Service App** obtains the processing result and updates the status of each CT-e into ORACLE database with the received result (cancelled / rejected). <br> 10) **Web Service App** ends its communication with CT-e Web Service. |
| Error sequences | E1) Problem found in validation Digital Certificate (invalid, revoked, overdue, prerequisites violation, CNPJ). <br> E2) Problem found in validation XML file (size, formatting, CT-e Web Service status). <br> E3) Problem found in establishing connection with CT-e Web Service (invalid / absent Brazilian federation Unit, version issues). <br> E4) Problem found in validating withdrawing Ct-e batch (invalid / absent Brazilian federation Unit, invalid digital signature, invalid CNPJ, inconsistency in XML information). <br> *Error sequences start at point 6 of main success sequences.* <br> 7. CT-e Web Service rejects CT-e batch and issues an error code. |

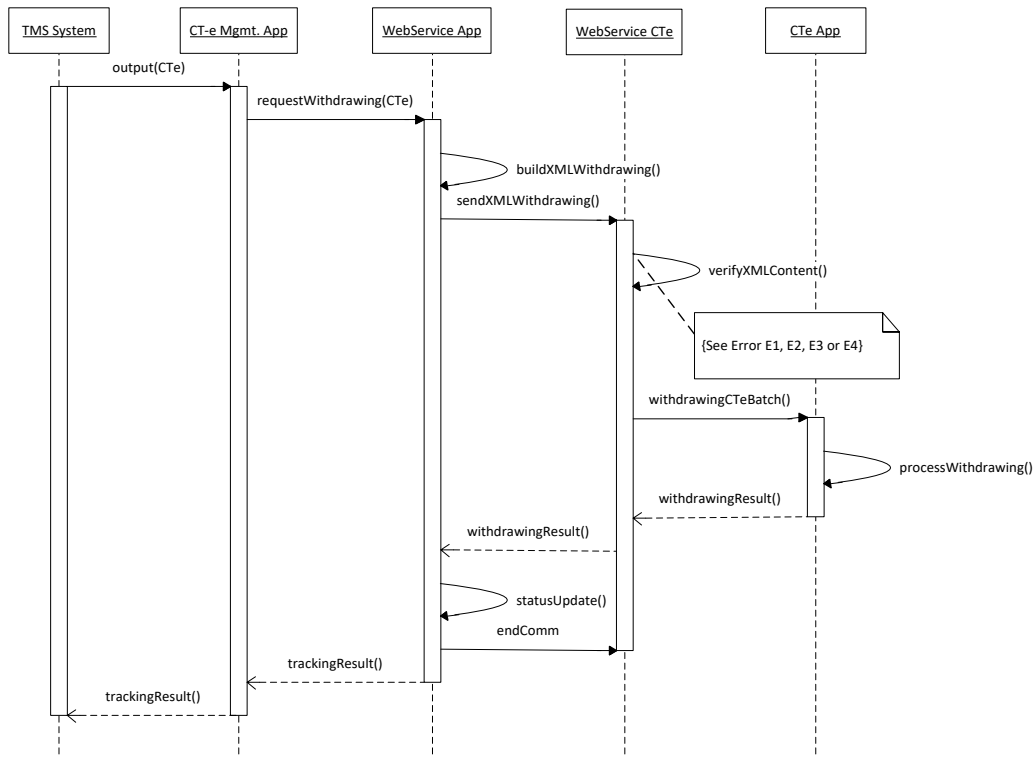

**Figure A.3. Detailed system sequence diagram of the *Withdrawing CT-e* main scenario.**

*A.4 Withdrawing CT-e numbering*

This use case allows the Web Service App, who is responsible to send CT-e batch data to the CT-e Web Service, to obtain a XML document from the CT-e Web Service containing the withdrawing processing result. This is a synchronous service provided by the CT-e Web Service responsible to withdraw CT-e numbering. As preconditions, it is required from Cargo carrier to be registered with SEFAZ, to have the digital certificate, a password issued by SEFAZ, a valid CT-e batch identifier (key). Table 5 presents textual description of this use case. Figure 4.5 illustrates the system sequence diagram.

**Table 4**
Textual description of Withdrawing CT-e numbering use case.

| Sequences | Description |
| --- | --- |
| Main Success sequences | 1) TMS system generates a TXT file containing all CT-e numbering to be withdrawn and sends it to **Windows Srv. App.**<br>2) **Windows Srv. App** obtains CT-e and sends to **Web Service App.**<br>3) **Web Service App** inserts CT-e into a queue up to 50 CT-e at a limit of 500k and then build a XML which consists of the CT-e batch.<br>4) **Web Service App** updates into ORACLE database CT-e status to "cancelling numbering".<br>5) **Web Service App** sends to CT-e Web Service each produced CT-e batch.<br>6) CT-e Web Service performs some initial validations (digital certificate, CT- batch size, XML formatting).<br>7) CT-e Web Service sends withdrawing request to CT-e Application.<br>8 ) CT-e Web Service issues a XML document containing the processing result.<br>9) **Web Service App** obtains the processing result and updates the status of each CT-e into ORACLE database with the received result (cancelled / rejected).<br>10) **Web Service App** ends its communication with CT-e Web Service. |
| Error sequences | E1) Problem found in validation Digital Certificate (invalid, revoked, overdue, prerequisites violation, CNPJ).<br>E2) Problem found in validation XML file (size, formatting, CT-e Web Service status).<br>E3) Problem found in establishing connection with CT-e Web Service (invalid / absent Brazilian federation Unit, version issues).<br>E4) Problem found in validating withdrawing Ct-e batch (invalid / absent Brazilian federation Unit, invalid digital signature, invalid CNPJ, inconsistency in XML information).<br>*Error sequences start at point 6 of main success sequences.*<br>7. CT-e Web Service rejects CT-e batch and issues an error code. |

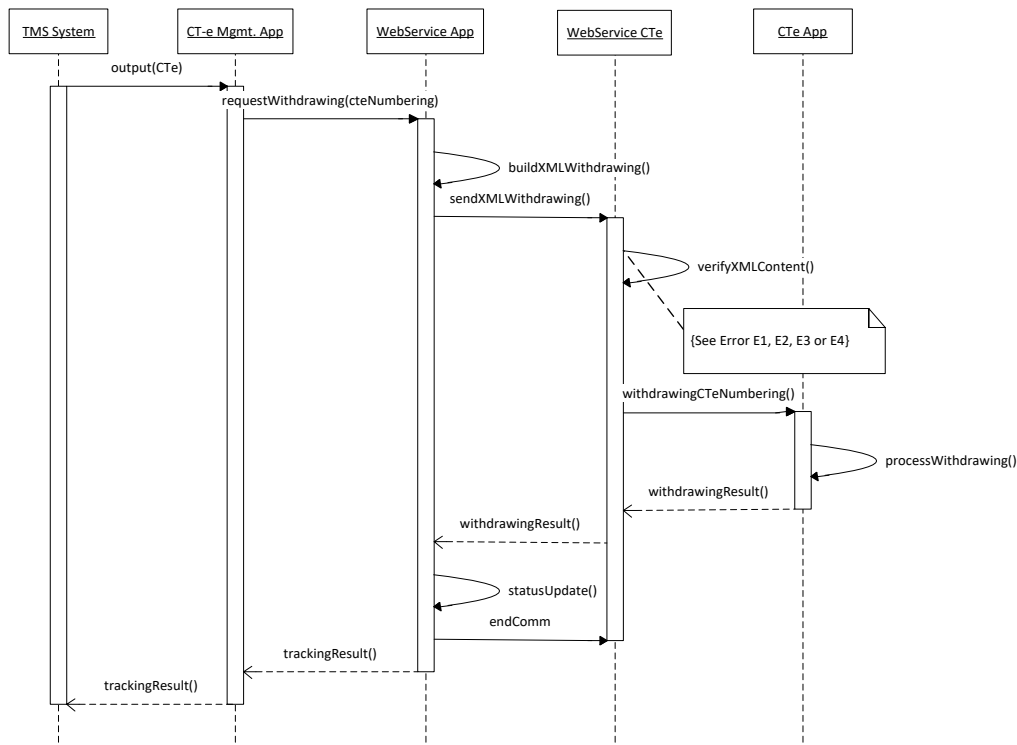

**Figure A.4.** Detailed system sequence diagram of the *Withdrawing CT-e Numbering* main scenario.

*A.5 Tracking CT-e current status*

This use case allows the Web Service App, who is responsible to send CT-e batch data to the CT-e Web Service, to obtain a XML document from the CT-e Web Service containing the processing result of each CT-e within a batch. This is a synchronous service provided by the CT-e Web Service responsible to track CT-e current status. As preconditions, it is required from Cargo carrier to be registered with SEFAZ, to have the digital certificate, a password issued by SEFAZ, a valid CT-e batch identifier (key). Table 6 presents textual description of this use case. Figure 4.6 illustrates the system sequence diagram.

**Table 5**
Textual description of Tracking CT-e current status use case.

| Sequences | Description |
|---|---|
| Main Success sequences | 1) Once the processing result of *Sending CT-e Batch* is available and favourable, **Web Service App** automatically builds a XML document comprising CT-e information to be tracked.<br>2) **Web Service App** sends the XML document to CT-e Web Service.<br>3) CT-e Web Service performs some initial validations (digital certificate, CT- batch size, XML formatting).<br>4) CT-e Web Service sends the tracking request to CT-e Application<br>5) CT-e Web Service issues a XML document containing the processing result.<br>6) **Web Service App** obtains the processing result and update in ORACLE database the "processing status" field of the CT-e with the received result (aproved / rejected).<br>7) **Web Service App** ends its communication with CT-e Web Service |
| Error sequences | E1) Problem found in validation Digital Certificate (invalid, revoked, overdue, prerequisites violation, CNPJ).<br>E2) Problem found in validation XML file (size, formatting, CT-e Web Service status).<br>E3) Problem found in establishing connection with CT-e Web Service (invalid / absent Brazilian federation Unit, version issues).<br>E4) Problem found in validating withdrawing Ct-e batch (invalid / absent Brazilian federation Unit, invalid digital signature, invalid CNPJ, inconsistency in XML information).<br>*Error sequences start at point 3 of main success sequences.*<br>4. CT-e Web Service rejects CT-e batch and issues an error code. |

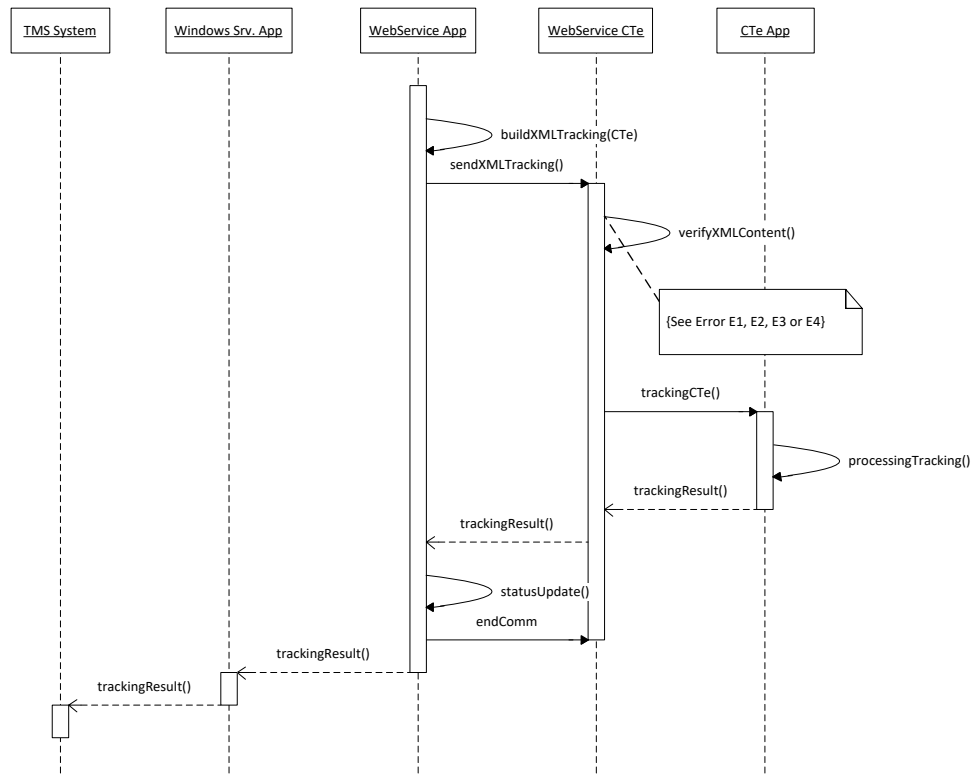

**Figure A.5** Detailed system sequence diagram of the *Tracking CT-e Current Status* main scenario.